\documentclass[11pt]{article}
\usepackage[dvips]{graphicx}
 \usepackage{amsmath}
\newcommand{\be}{\begin{displaymath}}
\newcommand{\bn}{\begin{equation}}
\newcommand{\bea}{\begin{eqnarray*}}
\newcommand{\eea}{\end{eqnarray*}}
\newcommand{\en}{\end{equation}}
\newcommand{\ee}{\end{displaymath}}

\newcommand{\lang}{\left\langle}
\newcommand{\rang}{\right\rangle}

\newcommand{\simlt}{\:{\raisebox{-1.5mm}{$\stackrel
{\textstyle{<}}{\sim}$}}\:}

\begin{document}

\begin{center}

\Large

{\bf Quasilinear particle transport from gyrokinetic instabilities in general magnetic geometry}
~\\
\normalsize

\vspace{1cm}
Per Helander and Alessandro Zocco
\\[1cm]
{\it Max-Planck-Institut f\"ur Plasmaphysik, 17491 Greifswald, Germany}

\vspace{0.5cm}

\end{center}

\noindent

The quasilinear particle flux arising from gyrokinetic instabilities is calculated in the electrostatic and collisionless approximation, keeping the geometry of the magnetic field arbitrary. In particular, the flux of electrons and heavy impurity ions is studied in the limit where the former move quickly, and the latter slowly, along the field compared with the mode frequency. Conclusions are drawn about how the particle fluxes of these species depend on the magnetic-field geometry, mode structure and frequency of the instability. Under some conditions, such as everywhere favourable or unfavourable magnetic curvature and modest temperature gradients, it is possible to make general statements about the fluxes independently of the details of the instability. In quasi-isodynamic stellarators with favourable bounce-averaged curvature for most particles, the particle flux is always outward if the temperature gradient is not too large, suggesting that it might be difficult to fuel such devices with gas puffing from the wall. In devices with predominantly unfavourable magnetic curvature, the particle flux can be inward, resulting in spontaneous density peaking in the centre of the plasma. In the limit of highly charged impurities, ordinary diffusion (proportional to the density gradient) dominates over other transport channels and the diffusion coefficient becomes independent of mass and charge. An estimate for the level of transport caused by magnetic-field fluctuations arising from ion-temperature-gradient instabilities is also given and is shown to be small compared with the electrostatic component. 

\noindent

\newpage

\section{Introduction}

Turbulence is ubiquitous in both tokamaks and stellarators, and generally degrades the energy confinement, which is therefore maximised if there is as little turbulence as possible in the plasma. Such a state 
is however undesirable because heavy impurity ions then tend to accumulate in the plasma core under the influence of neoclassical transport \cite{Taylor-1961,Connor-1973,Igitkhanov,Velasco,Helander-Newton,Newton}. This is a particular problem in stellarators, where neoclassical transport is much larger than in tokamaks. Moreover, in stellarators this transport is such as to cause a hollow electron density profile in the absence of turbulence, because the thermodiffusion -- i.e. particle transport driven by the temperature gradient -- is nearly always outward for the electrons. In the absence of a central particle source it thus becomes impossible to maintain a steady-state density profile \cite{Simmet}. Turbulence is however usually unavoidable, and it is therefore of interest to try to predict its effect on the particle transport. In tokamaks, this has been done with considerable success, using quasilinear theory of gyrokinetic instabilities \cite{Estrada-Mila,Angioni-1,Angioni-2,Garbet}, but no similar study of stellarator plasmas has been published, except for the study of impurity transport by Mikkelsen et al.~\cite{Mikkelsen}. 

This is the aim of the present article, where the gyrokinetic turbulent particle transport is calculated in the quasilinear approximation. As always in gyrokinetics, this transport is intrinsically ambipolar \cite{Sugama-1996}, so in a pure plasma the calculation can be done either for the ions or for the electrons; the two results always coincide. Of course, the gyrokinetic equation for both species needs to be solved to obtain the mode structure and frequency, but valuable information about the particle transport can be extracted from the electron equation alone. In contrast, little can be said using only the ion equation, for if the electrons for instance are taken to be adiabatic (Boltzmann-distributed), the particle flux vanishes identically. 

In an impure plasma, it is of great interest to calculate the transport of the impurity ions, which tend to spoil the energy confinement by causing excessive radiation losses. This transport can be calculated relatively easily for impurities that are sufficiently heavy that their motion along the magnetic field can be neglected in the gyrokinetic equation. 

The outcome of these calculations is a particle transport law for each species $a$,
	\bn \Gamma_a
	= - n_a \left( D_{a1} \frac{d \ln n_a}{d \psi} + D_{a2} \frac{d \ln T_a}{d \psi} + C_a \right),
	\label{transport law}
	\en
relating the cross-field particle flux $\Gamma_a$ to the density and temperature gradients as well as a term, here denoted by $C_a$, related to the curvature of the magnetic field. In tokamaks, this term usually describes inward-directed transport and is referred to as the ``curvature pinch''. 

In the following sections, we show how to calculate these transport coefficients for electrons and for highly charged impurity ions from gyrokinetic theory in the electrostatic and collisionless approximation. We discuss in detail how the outcome depends on the geometry of the magnetic field and leads to qualitative differences between different types of magnetic confinement devices. We also briefly assess the importance of electromagnetic effects on the transport caused by ion-temperature-gradient instabilities. 

\section{Linear gyrokinetics}

Our analysis proceeds from the linearised gyrokinetic equation in the electrostatic and collisionlesss approximation. The notation and the analysis in this section follow Ref.~\cite{Helander-TEM}, but are reproduced here for convenience. The distribution function of each species $a$ is written as
	$$ f_{a} = f_{a0}({\bf r},v) \left(1 - \frac{e_a \phi ({\bf r},t)}{T_a} \right) 
	+ g_a({\bf R}, v, \lambda, t), $$
where $\bf r$ denotes the particle position, $\bf R$ the guiding-centre position, $\phi$ the electrostatic potential, $e_a$ the charge, $v$ the speed, $\lambda = v_\perp^2/(v^2 B)$ the ratio between the magnetic moment $\mu = m_a v_\perp^2/(2B)$ and the kinetic energy $m_av^2/2 = x^2 T_a$, and 
	$$ f_{a0} = n_a \left( \frac{m_a}{2\pi T_a} \right)^{3/2} e^{-x^2}$$
the Maxwellian with number density $n_a(\psi)$ and temperature $T_a(\psi)$. The magnetic field is written in Clebsch form, ${\bf B} = \nabla \psi \times \nabla \alpha$, the wave vector in ballooning space as ${\bf k}_\perp = k_\psi \nabla \psi + k_\alpha \nabla \alpha$, and the diamagnetic and drift frequencies, respectively, are defined by 
	$$ \omega_{\ast a} = \frac{k_\alpha T_a}{e_a} \frac{d \ln n_a}{d \psi}, $$
	$$ \omega_{\ast a}^T = \omega_{\ast a} \left[ 1 + \eta_a \left( x^2 - \frac{3}{2} \right) \right], $$
	$$ \omega_{da} = {\bf k}_\perp \cdot {\bf v}_{da}, $$
where $\eta_a = d \ln T_a / d \ln n_a$ and ${\bf v}_{da}$ denotes the drift velocity. In this notation, the equation for $g_a$ becomes in Fourier (ballooning) space
	\begin{equation}
	i v_{\|}\nabla_{\|}g_a+(\omega -\omega_{da})g_a 
	= J_0 \left(\frac{k_\perp v_\perp}{\Omega_a} \right) \frac{e_a\phi}{T_a} \left( \omega - \omega_{*a}^T 
	\right)f_{a0}, 
	\label{gk}
	\end{equation}
where $\Omega_a = e_a B / m_a$ the gyrofrequency and the derivatives are taken at fixed energy and magnetic moment. $J_0$ is the zeroth-order Bessel function of the first kind, and corresponds to a gyroaveraging operator in real space. 

The gyrokinetic equation (\ref{gk}) usually needs to be solved numerically. Indeed, great efforts have gone into the construction of computer codes for this purpose. As is well known, it is however a simple matter to solve the equation analytically in the limit of fast and slow-moving particles. 

For gyrokinetic instabilities other than the electron-temperature-gradient mode, the wavelength somewhat exeecs the ion gyroradius, $k_\perp \rho_i \simlt 1$, and the frequency is of order 
	\bn \omega \sim \omega_{\ast i} \sim \frac{k_\perp \rho_i v_{Ti}}{L_\perp}, 
	\label{omega estimate}
	\en
where the subscript $i$ refers to the bulk ions and $L_\perp$ denotes the scale length of the density and temperature gradients. This frequency therefore lies well below the electron bounce/transit frequency, $\omega_{be} \sim v_{Te} / L_{\|}$, the inverse time it typically takes for a thermal electron to move the distance $L_\|$ between two bounce points or stellarator modules. The electron distribution function can then be expanded in the small parameter 
	\bn \frac{\omega}{\omega_{be}} \sim k_\perp \rho_i \sqrt{\frac{m_e}{m_i}} \; \frac{L_\|}{L_\perp} \ll 1, 
	\label{ordering}
	\en
and we thus write $g_e = g_{e0} + g_{e1} + \cdots$ with
	$$ v_\| \nabla_\| g_{e0} = 0. $$
This equation implies that $g_{e0}$ vanishes for untrapped particles, since it must do so at infinity in ballooning space, whereas for trapped ones
	\bn 
	i v_{\|}\nabla_{\|}g_{e1}+(\omega -\omega_{de})g_{e0} 
	= -\frac{e  J_0 \phi}{T_e} \left( \omega - \omega_{\ast e}^T 
	\right)f_{e0}. 
	\label{gk1}
	\en
In each magnetic well, labelled by an integer $j$ and defined by the condition $\lambda B = (v_\perp/v)^2 < 1$, the first term in this equation is annihilated by a bounce average, 	
	$$ \overline{\phi}_j(\lambda) 
	= \frac{1}{\tau_j} 
	\int \frac{\phi(l) \; dl}{\sqrt{1 - \lambda B(l)}} , $$
where the integral is carried out between the two bounce points of the $j$:th well, $l$ denotes the arc length along the magnetic field, and
	\bn \tau_j(\lambda) = \int \frac{dl}{\sqrt{1 - \lambda B(l)}} 
	\label{tau}
	\en
the normalised bounce time. The application of this bounce average to Eq.~(\ref{gk1}) gives
	\bn g_{e0}^{\rm tr} = - \frac{\omega - \omega_{*e}^T}{\omega - \overline{\omega}_{de}} 
	\frac{e \overline{J_0 \phi}_j}{T_e} \; f_{eo} 
	\label{ge}
	\en
in each trapping region. Since $g_e$ vanishes in leading order for circulating particles but not for trapped ones, we can already conclude that most of the transport is carried by the latter. The physical reason is that, in the ordering $\omega \ll \omega_{be}$, circulating electrons travel many turns around the torus on the time scale of the instability and therefore only experience a small average perturbation field. (In practice, some numerical studies have however indicated that in practice $\omega/\omega_{be}$ may be large enough to cause some significant transport of circulating electrons \cite{Hallatschek,Jenko}.)

The opposite limit, where the mode frequency exceeds the bounce frequency, applies to heavy ions, particularly impurities with charge numbers $Z \sim m_Z/m_i \gg 1$. Specifically, referring to the ordering (\ref{ordering}), we take
	\bn \frac{\omega}{\omega_{bZ}} \sim k_\perp \rho_i \sqrt{Z} \; \frac{L_\|}{L_\perp} \gg 1.
	\label{impurity ordering}
	\en
It then follows that
	$$ \frac{\omega}{\omega_{dZ}} \sim \frac{L_\| Z}{L_\perp} \gg 1, $$ 
and the solution of the gyrokinetic equation (\ref{gk}) becomes to leading order
	\bn g_Z = \left(1 - \frac{\omega_{\ast Z}^T}{\omega} \right) J_0 \frac{Ze\phi}{T_Z} f_{Z0}. 
	\label{gZ}
	\en
	
\section{Quasilinear particle flux}

We now set out to calculate the particle flux of electrons and impurity ions, respectively. We do so using the quasilinear approximation, in which the flux is calculated from the linear solutions (\ref{ge}) and (\ref{gZ}), respectively, of the linear gyrokinetic equation.  

To find the flux, we first need the cross-field ${\bf E} \times {\bf B}$ velocity, 
	$$ {\bf v}_E \cdot \nabla \psi = \frac{{\bf b} \times J_0 \nabla \phi}{B} \cdot \nabla \psi = - i k_\alpha J_0 \phi, $$
where $J_0$ again denotes the gyro-averaging operator. The quasilinear turbulent flux of any species $a$ is thus given by
	$$ \Gamma_a = \Re \lang \int f_a {\bf v}_E^\ast \cdot \nabla \psi \; d^3v \rang 
	= -  k_\alpha \Im \lang \int g_a J_0 \phi^\ast d^3v \rang, $$
where an asterisk indicates complex conjugation and angular brackets the flux-surface average \cite{Helander-ROP}
	$$ \lang \cdots \rang = \lim_{L \rightarrow \infty} \int_{-L}^L (\cdots) \frac{dl}{B(l)} 
	\bigg\slash \int_{-L}^L \frac{dl}{B(l)}. $$
	
\subsection{Electron flux}
	
Using Eq.~(\ref{ge}) we thus obtain the electron flux as
	$$ \Gamma_e = k_\alpha  \Im \lang \phi^\ast \int_{\rm tr.} 
	\frac{\omega - \omega_{*e}^T}{\omega - \overline{\omega}_{de}} 
	J_0 \frac{e \overline{J_0 \phi}_j}{T_e} \; f_{eo} d^3v  \rang. $$
Here, the integral is only taken over the trapped part of velocity space, and can be simplified by using 
	$$ \int_{-\infty}^\infty \phi^\ast(l) dl \int_{1/B_{\rm max}}^{1/B_{\rm min}} \frac{\overline{\phi}_j \; d\lambda}{\sqrt{1- \lambda B}}
	= \int_{1/B_{\rm max}}^{1/B_{\rm min}} d\lambda \sum_j \tau_j | \overline{\phi}_j |^2, $$
where the sum is taken over all relevant magnetic wells (regions with magnetic field strength $B < 1/\lambda$). The electron flux can thus be written as
	$$ \Gamma_e = \frac{e k_\alpha }{T_e} \Im \int_0^\infty f_{e0} 2 \pi v^2 dv
	\int_{\rm tr.} 	\sum_j \frac{\omega - \omega_{*e}^T}{\omega - \overline{\omega}_{de}} 
	\tau_j |\overline{J_0 \phi}_j |^2 d\lambda \bigg\slash \int \frac{dl}{B}, $$
where the $\lambda$-integral is taken over all trapped particles, i.e. 
	$$ \frac{1}{B_{\rm \max}} < \lambda < \frac{1}{B_{\rm min}}, $$
with $B_{\rm min}$ and $B_{\rm max}$ denoting the minimum and maximum field strength on the flux surface in question. This expression for the flux can be simplified by using Eq.~(\ref{tau}) and interchanging the integrations in $l$ and $\lambda$, giving
	$$ \Gamma_e = \frac{e k_\alpha }{T_e} 
	\Im \lang \int_{\rm tr.} \frac{\omega - \omega_{*e}^T}{\omega - \overline{\omega}_{de}}
	f_{e0} |\overline{J_0 \phi} |^2 d^3v \rang. $$
When interpreting this expression, it must be remembered that the quantities $\overline{\omega}_{de}$ and $\overline \phi$ depend both on the velocity space variable $\lambda$ (through the bounce average) and, discretely, on the coordinate along the field line $l$, which selects the appropriate trapping well over which the bounce average is taken. 

In quasilinear theory, only linearly growing modes produce transport, damped ones do not. We therefore write $\omega = \omega_r + i \gamma$ with $\gamma > 0$, and define the function
	$$ \Delta_\gamma (x) = \frac{\gamma/\pi}{\gamma^2 + x^2}, $$
which approaches a Dirac delta function in the limit $\gamma \rightarrow 0+$. In this notation, we can write the electron flux as
	\bn \Gamma_e \frac{d \ln n_e}{d \psi} 
	= - \pi \omega_{\ast e} \lang \int_{\rm tr.} \left| \frac{e \overline{J_0 \phi}}{T_e} \right|^2
	\Delta_\gamma (\omega_r - \overline{\omega}_{de})
	\left( \omega_{*e}^T - \overline{\omega}_{de} \right) f_{e0} d^3v \rang
	\label{e-flux}
	\en
	$$ = - \frac{d n_e}{d \psi}\left( D_{e1} \frac{d \ln n_e}{d \psi} + D_{e2} \frac{d \ln T_e}{d \psi} + C_e
	\right),
	$$
where we have recalled Eq.~(\ref{transport law}) and defined
	$$ D_{e1} = \frac{\pi k_\alpha^2}{n_e} \lang \int_{\rm tr.} \left| \overline{J_0 \phi} \right|^2 
	\Delta_\gamma (\omega_r - \overline{\omega}_{de}) f_{e0} d^3v \rang, $$
	$$ D_{e2} = \frac{\pi k_\alpha^2}{n_e} \lang \int_{\rm tr.} \left| \overline{J_0 \phi} \right|^2 
	\left(x^2 - \frac{3}{2} \right) \Delta_\gamma (\omega_r - \overline{\omega}_{de}) f_{e0} d^3v \rang, $$
	\bn C_e = \frac{\pi e k_\alpha}{n_e T_e} \lang \int_{\rm tr.} \left| \overline{J_0 \phi} \right|^2 
	\Delta_\gamma (\omega_r - \overline{\omega}_{de}) \overline{\omega}_{de} f_{e0} d^3v \rang. 
	\label{Ce}
	\en

We thus see that the particle flux is a sum of three terms, proportional to the density and temperature gradients in $\omega_{\ast e}^T$ and to the magnetic curvature in $\overline \omega_{de}$, respectively. Since $\gamma$ must be positive for quasilinear theory to apply, the function $\Delta_\gamma$	is also positive, and it follows that the diffusion coefficient multiplying the density gradient is always positive, $D_{e1} > 0$, as required by the condition of positive entropy production. (Quasilinear transport satisfies both an H-theorem and Onsager symmetry.) The term containing the temperature gradient $dT_e/d\psi$ is called thermodiffusion and can have either sign. The term $C_e$ from the magnetic curvature is usually called the `curvature pinch' on the grounds that it usually describes inward transport in tokamaks, but as we shall see it can be directed outward in stellarators.  
	
\subsection{Impurity ion flux}

Similarly, but somewhat more straightforwardly, the quasilinear flux of heavy impurity ions can be obtained from Eq.~(\ref{gZ}),
	$$ \Gamma_Z = - \frac{Z e k_\alpha }{T_Z} \Im \lang \int 
	\left( 1 - \frac{\omega_{\ast Z}^T}{\omega} \right) |J_0 \phi |^2  f_{Z0} d^3v \rang, $$
and has the same form as that for the electrons
	\bn \Gamma_Z 
	= - n_Z \left( D_{Z1} \frac{d \ln n_Z}{d \psi} + D_{Z2} \frac{d \ln T_Z}{d \psi} + C_Z
	\right).
	\label{Z-flux}
	\en
In the approximation (\ref{gZ}), the transport coefficients can be calculated explicitly and are equal to 
	\bn D_{Z1} = \frac{\gamma k_\alpha^2}{\omega_r^2 + \gamma^2} \lang |\phi|^2 \Gamma_0(b) \rang, \label{DZ1} \en
	\bn D_{Z2} = \frac{\gamma k_\alpha^2}{\omega_r^2 + \gamma^2} \lang |\phi|^2 b \left[\Gamma_1(b) - \Gamma_0(b) \right] \rang, \label{DZ2} \en
	$$ C_Z = 0, $$
where $b = k_\perp^2 T_Z / (m_Z \Omega_Z^2)$, $\Gamma_n(b) = I_n(b) e^{-b}$ and $I_n$ is the modified Bessel function of order $n$.  
Note that, in constrast to the electrons, not only trapped ions but also circulating ones contribute to the transport. Again, $D_{Z1}$ is positive whereas $D_{Z2}$ is always negative in this approximation.  

\section{Discussion of the results}

The expressions (\ref{e-flux}) and (\ref{Z-flux}) are generally valid within the assumptions of the model, and we now turn to a detailed discussion of these results. The exact calculation of the fluxes would require knowledge of the eigenfunction $\phi(l)$ and the eigenvalue $\omega$ of the linear stability problem -- information that is only available numerically from gyrokinetic codes. However, even without such information, it is possible to draw interesting conclusions about the nature and the direction of the transport in a number of important special cases and in certain mathematical limits of the plasma parameters.

\subsection{Impurity transport}

The result (\ref{Z-flux}) for the impurity transport is much simpler than its electron counterpart (\ref{e-flux}) and becomes yet simpler if one notices that, in the limit of heavy impurities, the impurity gyroradius becomes much smaller than that of the ions, since $\rho_Z / \rho_i \sim Z^{-1/2}$, so we expect $b \ll 1$ and $D_{Z2} \ll D_{Z1}$. To a first approximation, there is thus neither thermodiffusion nor a curvature pinch for heavy impurities. Moreover, the remaining diffusion coefficient (\ref{DZ1}) reduces to 
	$$ D_{Z1} \simeq \frac{\gamma }{\omega_r^2 + \gamma^2} \lang | k_\alpha \phi|^2 \rang $$
and thus becomes independent of both mass and charge, reflecting the corresponding property of the ${\bf E} \times {\bf B}$ drift in the limit of zero gyroradius. 

Small corrections to this result arises if one either includes finite-gyroradius effects coming from the Bessel function in Eq.~(\ref{gZ}), as was done in Eq.~(\ref{Z-flux}), or if the gyrokinetic equation is solved to higher accuracy and one accounts for the drift frequency $\omega_{dZ}$ by replacing (\ref{gZ}) by the slightly more accurate expression 
	\bn g_Z = \frac{\omega - \omega_{\ast Z}^T}{\omega - \omega_{dZ}} \;  J_0 \frac{Ze\phi}{T_Z} f_{Z0}. 
	\label{gZ1}
	\en
In analogy with Eq.~(\ref{e-flux}), the impurity flux then becomes
	\bn \Gamma_Z \frac{d\ln n_Z}{d\psi} 
	= - \pi \omega_{\ast Z} \lang \left| \frac{Ze\phi}{T_Z} \right|^2 \int \Delta_\gamma (\omega_r - \omega_{dZ})
	\left( \omega_{*Z}^T - \omega_{dZ} \right) J_0^2 f_{Z0} d^3v \rang,
	\label{Z-flux1}
	\en
and contains both finite thermodiffusion and a curvature pinch. 

\subsection{Good average curvature}

One case in which something definite can be said about the electron transport occurs when the bounce-averaged magnetic curvature is favourable (or `good') for all orbits, i.e. $\omega_{\ast e} \overline \omega_{de} < 0$, and, additionally, $\eta_e$ lies between 0 and 2/3. These conditions, which can be satisfied in quasi-isodynamic stellarators \cite{Proll-PoP}, also imply that collisionless trapped-electron modes are stable, both lineraly \cite{Proll} and nonlinearly \cite{Helander-2017}. When these two conditions are satisfied, then $ \omega_{\ast e}^T \overline \omega_{de} $ is negative throughout velocity space and according to Eq.~(\ref{e-flux}), where the function $\Delta_\gamma(\omega_r - \overline{\omega}_{de})$ is always positive, we have
	$$ \Gamma_e \frac{d \ln n_e}{d \psi} < 0, $$
so that the electron particle flux is in the direction of lower density.  

\subsection{Bad magnetic curvature}

If the magnetic curvature is everywhere unfavourable, $\omega_{\ast a} \omega_{da} > 0$, and the temperature gradient is neither too large nor too small, definite statements can be made about the direction of both the electron and impurity ion fluxes, as can be seen from the expression
	\bn \omega_{\ast a} \left( \omega_{*a}^T - {\omega}_{da} \right)
	= \omega_{\ast a}^2 \left[ 1 - \frac{3 \eta_a}{2} 
	+ x^2 \left( \eta_a - \frac{\tilde \omega_{da}}{\omega_{\ast a}} \right) \right], 
	\label{bad-curvature case}
	\en
where we have written $\omega_{da} = \tilde \omega_{da}(\lambda) x^2$, noting that the drift velocity is proportional to energy. If now
	\bn \frac{2}{3} < \eta < \min_\lambda \frac{\tilde \omega_{da}(\lambda)}{\omega_{\ast a}}, 
	\label{condition}
	\en
the expression (\ref{bad-curvature case}) is negative definite and it follows from Eqs.~(\ref{e-flux}) and (\ref{Z-flux1}) that
	$$ \Gamma_e \frac{d \ln n_e}{d \psi} > 0, $$
and
	$$ \Gamma_Z \frac{d \ln n_Z}{d \psi} > 0, $$
so both the electron and impurity fluxes are in the direction of the respective density gradient, i.e. usually into the plasma. The resulting density peaking in the centre of the plasma is generally desirable for the electrons but undesirable for the impurities. Note that the condition (\ref{condition}) depends on the density gradient for the particle species in question and therefore gets progressively more difficult to satisfy when the density profile becomes steeper. The peaking of this profile will stop when the particle flux balances the sources within the plasma. 

The condition that the magnetic curvature be unfavourable everywhere applies in a $Z$-pinch, a screw-pinch, a reversed-field pinch and in a dipole magnetic field. In a typical tokamak or stellarator, however, the magnetic curvature is good in some places on each magnetic surface and bad in others, but the fluctuation amplitude $| \phi |$ usually peaks in the bad-curvature region on the outboard side of the torus. If this peaking is sufficiently strong, the conclusions of this subsection then apply. In particular, if $\eta > 2/3$ and
the transport caused by the temperature gradient can then be so strong that it exceeds that from the density gradient, resulting in a density profile that peaks in the core of the plasma, regardless of other properties of the turbulence. 

\subsection{Thermodiffusion}

We now specifically consider the effect of the temperature gradient on the particle transport. From Eq.~(\ref{e-flux}) the electron thermodiffusion coefficient is equal to
	\bn D_{e2} = \frac{\gamma}{n_e} \left( \frac{k_\alpha T_e}{e} \right)^2
	\lang \int_{\rm tr.} \left| \overline \phi \right|^2
	\frac{x^2 - 3/2}{\gamma^2 + (\omega_r - \overline{\omega}_{de})^2} f_{e0} d^3v \rang,
	\label{De2}
	\en
where we have set the Bessel function equal to unity since wavelength of most relevant microinstabilities is much larger than the electron gyroradius. As Angioni et al.~\cite{Angioni-1} have pointed out, the integrand is positive for $x^2 > 3/2$ and negative for $x^2 < 3/2$, and were it not for the energy-dependence of $\overline{\omega}_{de}$ in the denominator, the integral would vanish. Ion-temperature-gradient (ITG) modes in tokamaks usually propagate in the ion diamagnetic direction, which we take to be positive, $\omega > 0$, and since $|\phi|$ peaks on the outboard side of the torus, $\overline{\omega}_{de}$ is negative, making $(\omega_r - \overline{\omega}_{de})^2$ an increasing function of energy. The expression (\ref{De2}) then becomes negative, implying inward thermodiffusion of electrons. If, on the other hand, $\omega$ and $\overline{\omega}_{de}$ have the same sign and $\omega / \overline{\omega}_{de} \gg 1$, as expected for trapped-electron modes (TEMs), then the integral in Eq.~(\ref{De2}) becomes negative, implying outward thermodiffusion. This is thought to explain `density pump-out' in tokamaks with electron cyclotron resonance heating \cite{Angioni-2}. 

A similar argument can now be made for thermodiffusion of impurity ions. According to Eq.~(\ref{Z-flux}), the thermodiffusion coefficient is 
	$$ D_{Z2} = \frac{\gamma}{n_Z} \left( \frac{k_\alpha T_Z}{Ze} \right)^2
	\lang \left|  \phi \right|^2  
	\int \frac{x^2 - 3/2}{\gamma^2 + (\omega_r - {\omega}_{dZ})^2} f_{Z0} d^3v \rang, $$
where we have ignored the Bessel function on the grounds that its argument is much smaller for heavy ions than for light ones. If it is of order unity for the bulk ions, it should thus be permissible to neglect finite-gyroradius effects for heavy impurities. For ITG modes in tokamaks, and for other modes propagating in the same direction of the impurity drift, $\omega \omega_{dZ} > 0$, we expect outward thermodiffusion, $D_{Z2} > 0$. On the other hand, if the Bessel function is not negligible, there is inward thermodiffusion according to Eq.~(\ref{DZ2}). 

\subsection{Curvature pinch}

The presence of a `curvature-pinch' term in Eqs.~(\ref{transport law}), (\ref{e-flux}) and (\ref{Z-flux1}) is fundamentally interesting since it implies that the state of zero particle flux is not necessarily one where the density and temperature gradients vanish. Even if they do, there is in general a finite particle flux, which reflects a tendency of the plasma to `spontaneously' develop a density gradient \cite{Garbet,Isichenko}. The underlying reason is that the lowest accessible energy state is generally not homogeneous if the plasma dynamics is contrained to conserve adiabatic invariants \cite{Helander-2017}. 

In the present context, the curvature pinch can be read off from Eqs.~(\ref{Ce}) and (\ref{Z-flux1}):
	$$ C_e = \frac{\pi e k_\alpha}{n T_e} \lang \int_{\rm tr.} 
	\left| \overline{J_0 \phi} \right|^2 \overline{\omega}_{de} 
	\Delta_\gamma (\omega_r - \overline{\omega}_{de})
	 f_{e0} d^3v \rang,
	$$
	$$ C_Z = - \frac{\pi Z e k_\alpha}{n_Z T_Z} \lang \int \left| J_0 \phi\right|^2 
	\omega_{dZ} 
	\Delta_\gamma (\omega_r - \omega_{dZ})
	 f_{Z0} d^3v \rang.
	$$
Hence it is clear that the curvature pinch is indeed caused by magnetic curvature through the drift frequency $\omega_d$. Its direction is most clearly understood from a comparison with ordinary density-gradient-driven diffusion,
	$$ \frac{C_e}{D_{e1} \frac{d \ln n}{d \psi}} = 
	- \lang \int_{\rm tr.} \frac{\overline{\omega}_{de}}{\omega_{\ast e}}  
	\left| \overline{ J_0 \phi} \right|^2
	\Delta_\gamma
	 f_{e0} d^3v \rang \bigg\slash
	\lang \int_{\rm tr.}  
	\left| \overline{ J_0 \phi} \right|^2
	\Delta_\gamma f_{e0} d^3v \rang. $$
Hence it is evident that the curvature pinch is indeed a `pinch', i.e. it transports electrons up the density gradient, if the bounce-averaged magnetic curvature is unfavourable, $\overline{\omega}_{de} \omega_{\ast e} > 0$, for most relevant orbits. This tends to be the case in tokamaks \cite{Garbet,Isichenko} but not in quasi-isodynamic stellarators \cite{Helander-TEM,Proll}, where the curvature pinch is thus outward \cite{Mishchenko}. 
	
Such devices thus suffer from a curvature `anti-pinch', potentially leading to hollow density profiles. However, the corresponding impurity flux, 
	$$ \frac{C_Z}{D_{Z1} \frac{d \ln n_Z}{d \psi}} =
	- \lang \int \frac{\omega_{dZ}}{\omega_{\ast Z}}  
	\left| J_0 \phi \right|^2
	\Delta_\gamma f_{Z0} d^3v \rang \bigg\slash
	\lang \int 
	\left| J_0 \phi \right|^2
	\Delta_\gamma f_{Z0} d^3v \rang, $$
is beneficial, driving the impurities in the same direction as ordinary diffusion, if the fluctuations peak in regions of favourable local magnetic curvature, $\omega_{dZ} \omega_{\ast Z} < 0$. 
	
\subsection{Near-marginality}

Just above the linear stability threshold, $\gamma \rightarrow 0+$, the function $\Delta_\gamma$ approaches a Dirac delta function, which simplifies the calculation of the velocity-space integral in Eqs.~(\ref{e-flux}) and (\ref{Z-flux}). One of the two velocity-space integrals can then be carried out, but the resulting expression does not yield much more information than already discussed above. 

For instance, the electron flux (\ref{e-flux}) becomes in this limit
	$$ \Gamma_e = - k_\alpha^2 \frac{dn}{d\psi} \int_{1/B_{\rm max}}^{1/B_{\rm min}}
	\sum_j \frac{\tau_j | \overline{J_0 \phi} |^2}{|\tilde \omega_{de}|} 
	\sqrt{\frac{\pi \omega}{\tilde \omega_{de}}} \; \Theta(\omega \tilde \omega_{de}) 
	e^{- \omega / \tilde \omega_{de}} $$
		$$ \times \left[1 + \eta_e \left( \frac{\omega}{\tilde \omega_{de}} - \frac{3}{2} \right) 
	- \frac{\omega}{\omega_{\ast e}} \right] d\lambda \bigg\slash \int \frac{dl}{B}, $$
where $\Theta$ denotes the Heaviside function, which in this expression ensures that the  frequencies $\omega$ and $\tilde \omega_{de}(\lambda)$ have the same sign, so that only resonant orbits contribute to the flux. From this result, we again see that the flux is in the direction of increasing density if the condition (\ref{condition}) is satisfied. We also note that thermodiffusion is outward if $\omega / \tilde \omega_{de} > 3/2$, but otherwise it is difficult to draw any general conclusions.

\section{Extension to electromagnetic instabilities}

The two main limitations of our results are the neglect of collisions and electromagnetic effects, which arise at finite plasma beta and perturb the equilibrium magnetic field. The electromagnetic fluctuation components are of two types: one associated with magnetic compressibility, $\delta B_\|$, and the other with ``magnetic flutter'', $\delta {\bf B}_\perp$ arising from perturbed electric currents parallel to ${\bf B}$. Magnetic compressibility affects ion-temperature-gradient modes if the electron pressure is of the order of $\beta_e \sim L_p /R$, where $L_p$ denotes the pressure gradient length scale and $R$ the curvature radius of the magnetic field \cite{ZHC}. As seen in gyrokinetic simulations, the destabilising effect of $\delta B_\|$ is largely cancelled by a reduction in the equilibrium $\nabla B$-drift (at constant magnetic curvature), so that little net effect is seen if one is careful to keep the magnetic drifts consistent with the magnetic equilibrium \cite{ZHC,Waltz,Joiner,Belli}. This is true even when $\beta$ is large enough to excite kinetic ballooning modes \cite{Aleynikova}.

Perpendicular magnetic fluctuations, coming from a term containing the magnetic potential $A_\|$ in the gyrokinetic equation, arise already at smaller pressures, of order \cite{ZHC}
	$$ \beta \sim \left( \frac{k_\perp \rho_i k_\| c_s}{\omega} \right)^2, $$
where $\rho_i$ denotes the ion gyroradius and $c_s$ the sound speed. These fluctuations cause extra particle transport by electrons streaming along perturbed magnetic field lines, but only if these reconnect and form overlapping magnetic islands, so that the field becomes chaotic. A quasilinear estimate of the island width, and of this transport, can be given if the electrostatic perturbation driven by ion-scale turbulence generates a magnetic perturbation at the electron scale, $\delta < \rho_i$, where magnetic flux surfaces are not preserved and reconnection can indeed occur. This is an intricate calculation that, to our knowledge, has only been performed in the case of the slab ion-temperature-gradient mode \cite{CHZ}. In  a more general setting, it is possible to highlight some features of the quasilinear stochastic transport without going into too many details. 
 First of all, the amplitude of the magnetic perturbation
that generates a magnetic island must be related to the amplitude
of the electrostatic potential, $\phi$. The amplitude of  electromagnetic fluctuations,
$A_{\parallel}$, can be considered the unknown of the ion-scale
problem  if it is assumed
that the solution of the leading-order electrostatic problem at the characteristic scale of the instability, $k_\psi \nabla \psi\sim k_{\alpha}/a$, is known, where $a \sim |\nabla \alpha |^{-1}$ denotes a normalising length. This means that the eigenfunction $\phi$
and the eigenvalue $\omega$ of the electrostatic mode (typically an ITG or TEM instability)
must be given for a specific wave number $k_{\alpha}/a$. Then, $A_{\parallel}$  is determined by multiple asymptotic matching between
the ion-region solution, $(k_{\psi}\nabla\psi) \rho_{i}\sim1,$ and the
leading-order electrostatic solution. At the
ion scale, $A_{\parallel}$ is a $\beta-$correction to the
electrostatic potential associated with the driving turbulence so
that $A_{\parallel}\text{\ensuremath{\propto}}\beta\phi,$
{[}see Eq. (11) of \cite{CHZ}{]}. Not surprisingly, this implies that
the addition of an electromagnetic component to Eq. \eqref{e-flux} would only contribute significantly to the transport if $\beta$ is large enough. We note that
the explicit form of $A_{\parallel}$ depends on the details
of the leading order electrostatic problem. Under the assumption
$\delta\ll\rho_{i}$, the matching of $A_{\parallel}$ found by perturbing the electrostatic problem to the value $\left.A_{\parallel}\right|_{\delta}$
evaluated from Ampere's law at the electron scale $\delta$, gives
the amplitude $\left.A_{\parallel}\right|_{\delta}$ as a function
of $\phi$. The specific form of the electron response is thus eventually required. In this regard, the analysis of Connor et al.~\cite{CHZ} was performed
for semicollisional electrons. However, a simple adaptation of the
their analysis to the collisionless case is straightforward and gives the magnetic perturbation at the rational surface
\begin{equation}
  v_{thi}A_{\parallel}(0)=\beta_{i}b_{0}^{2}\frac{\delta L_{s}}{\rho_{i} L_{T}}F(\frac{\omega_{0}}{\eta_e\omega_{*e}},\beta_{i}\frac{\rho_iL_{s}^{2}}{\delta L_{T}^{2}},b_{0},I_{e}) \phi,
\label{eq:aparzero}
\end{equation}
where the function $F$ takes into account the details of the driving
instability at the ion scale (i.e. it is a function of the $\Delta^{*}$
quantity introduced by the authors of Ref.~\cite{CHZ}), and given by the electrostatic problem, $I_{e}$ is an $\mathcal{O}(1)$
number resulting from the integration of a function of the collisionless
electron conductivity \cite{Drake-Lee,CHZ1,ZMicroTearing}, $b_0=k_{\alpha}^{2}(\rho_i/a)^{2}/2,$ 
$L_{T}^{-1}=d\log T/d\hat{\psi}$, $L_s$ denotes the shear length, $\hat{\psi}=\psi/(a^{2}B_0),$ and $B_0$ a reference magnetic field. In Eq. \eqref{eq:aparzero}, the explicit dependence on 
$b_0$ and $L_s/L_T$ are specific to the slab ITG mode, however the $\delta/\rho_i$ and $\beta_i-$dependencies are general. 
By using $\delta B_{r}\sim (k_{\alpha}\nabla\alpha) A_{\parallel}(0),$
a quasilinear estimate for the collisionless Rechester-Rosenbluth
transport coefficient gives \cite{Rech-Rosen}
\begin{equation}
  \chi_{e}= v_{the} L \left( \frac{\delta B_\perp}{B} \right)^2 \sim \left(\frac{v_{the}}{k_{\alpha}\left|\nabla\alpha\right|}\right)\frac{\beta_{i}^{2}b_{0}^{2}}{T_{e}^{2}/T_{i}^{2}}\frac{\delta L_{s}^{3}}{\rho_{i}^{2}L_{T}^{2}}F^{2}\frac{k_{\alpha}^{2}\left|\nabla\alpha\right|^{2}\phi^{2}}{v_{thi}^{2}B_{0}^{2}},
\end{equation}
where $L=L_{s}/(k_{\alpha}\left|\nabla\alpha\right|\delta)$ is the length that a particle needs to travel along the magnetic field to encounter the electron scale $\delta$. Let us now consider the particle transport due to the effective
diffusion caused by magnetic perturbations: $\partial_{t}n=\chi_{e}\nabla^{2}n.$
This will compete with the radial $\mathbf{E}\times\mathbf{B}$ transport
of Eq. \eqref{e-flux}: $\Gamma_{e}d\ln n/d\psi=\omega_{*e}nI_{p},$ where
$I_{p}$ is an $\mathcal{O}(1)$ quantity defined by the right-hand side of Eq.~\eqref{e-flux}
. We can now compare the ratio of the particle diffusion due to the stochastic
magnetic field, and the one given by the radial component of the $\mathbf{E}\times\mbox{\ensuremath{\mathbf{B}}}$
drift
\begin{equation}
\begin{split} & \frac{\chi_{e}L_T^{-2}n}{\Gamma_{e}\frac{d}{d\psi}\ln n}=2a\nabla\alpha\frac{\beta_{i}^{2}b_{0}^{4}}{\left(T_{e}/T_{i}\right)^{5/2}}\sqrt{\frac{m_{i}}{m_{e}}}\frac{\delta L_{s}^{3}}{\rho_{i} L_{T}^{3}}\frac{F^{2}}{I_{p}}\frac{k_{\alpha}^{2}\left|\nabla\alpha\right|^{2}\phi^{2}}{v_{thi}^{2}B_{0}^{2}}\\
 & \sim\beta_{i}^{2}b_{0}^{4}\sqrt{\frac{T_i m_{i}}{T_e m_{e}}}  \frac{\delta}{\rho_{i}} \left(\frac{L_{s}}{L_{T}}\right)^{3}\ll1.
\end{split}
\end{equation}
Since $\delta < \rho_i$, and in accordance with the result of Connor et al.~\cite{CHZ}, this is a very small
number, whose smallness is dictated by the electron-ion scale separation, by $\beta,$ and by $b_0.$

We thus conclude that, at least in pure ITG turbulence, no significant contribution to the transport from field-line flutter is to be expected. This conclusion may change, of course, if other microinstabilities such as micro-tearing modes are excited \cite{Nevins}.

\section{Conclusions}

Our conclusions can be summarised as follows. Not all of them are novel, in particular those relating to tokamaks, but we nevertheless list all significant findings here for convenience:

\begin{itemize}

\item Most of the electron transport is carried by trapped particles. 

\item Highly charged impurities are mostly subject to ordinary Ficksian diffusion, proportional to the density gradient, with a diffusion coefficient that is approximately independent of mass and change. The transport from thermodiffusion and the curvature pinch is relatively small. 

\item In tokamaks, most trapped particles experience unfavourable magnetic curvature on an orbit-average. Instabilities propagating in the ion diamagnetic direction therefore tend to produce inward thermodiffusion, and those propagating in the electron direction cause outward thermodiffusion, which is thought to explain `density pump-out' during electron-cyclotron-resonance heating. The curvature pinch is directed inward and tends, in combination with ITG-driven thermodiffusion, to cause slightly peaked density profiles. 

\item In so-called maximum-$J$ devices \cite{Rosenbluth}, such as quasi-isodynamic stellarators \cite{Proll}, where most trapped electrons experience favourable magnetic curvature on a time average, the net particle transport is always in the direction of decreasing density if $0 < \eta_e < 2/3$, and the curvature pinch is outward. This result suggests that it could be relatively difficult to achieve efficient fuelling of such stellarators using gas puffing at the plasma edge. Such difficulties have indeed been observed in the first operational phase of Wendelstein 7-X. 

\item In magnetic configurations with unfavourable magnetic curvature everywhere, such as magnetic dipoles, reversed-field pinches, Z-pinches and screw pinches, both the electron and the impurity fluxes are in the direction of increasing density, i.e., usually into the plasma, if the condition (\ref{condition}) is satisfied. Spontaneous density peaking should then occur at least to the point where Eq.~(\ref{condition}) is no longer satisfied. 

\item In turbulence driven by the ion-temperature-gradient instability, the contribution to transport from parallel streaming along chaotic magnetic field lines is relatively small. 

\end{itemize}

The main limitation of our calculation is the neglect of collisions, which in particular for heavy impurities restricts its validity to high-temperature plasmas.

%%%%%%%%%%%%%%%%%
%%%%%%%%%%%%%%%%%%%%%%%%%%%%%%%%%%%%%%%%%%%%%%%%%%%%%%%%%%%%%%%%%%%%%%

\end{document}